\documentclass[twocolumn,superscriptaddress,showpacs,floatfix,amsmath,amssymb]{revtex4-1}
\usepackage{color}
\usepackage{graphicx}
\usepackage[dvipdfmx,colorlinks=true,linkcolor=blue,citecolor=blue,urlcolor=blue]{hyperref}
\usepackage[utf8]{inputenc}
\usepackage{float}

\newcommand{\bLi}{${}^7$Li}
\newcommand{\fLi}{${}^6$Li}
\newcommand{\bYb}{${}^{174}$Yb}
\newcommand{\bEr}{${}^{168}$Er}
\newcommand{\fEr}{${}^{167}$Er}
\newcommand{\bbEr}{${}^{166}$Er}

\newcommand{\micron}{\mu\mathrm{m}}
\newcommand{\nm}{\mathrm{nm}}
\newcommand{\ms}{\mathrm{ms}}
\newcommand{\s}{\mathrm{s}}
\newcommand{\G}{\mathrm{G}}
\newcommand{\mG}{\mathrm{mG}}
\newcommand{\Hz}{\mathrm{Hz}}
\newcommand{\kHz}{\mathrm{kHz}}

\newcommand{\uK}{\mu\mathrm{K}}

\begin{document}

\title{Feshbach resonances of large mass-imbalance Er-Li mixtures}

\author{F.~Sch\"{a}fer}
\email{schaefer@scphys.kyoto-u.ac.jp}
\affiliation{Department of Physics, Graduate School of Science, Kyoto University, Kyoto 606-8502, Japan}

\author{N.~Mizukami}
\altaffiliation[Present address: ]{European Laboratory for Nonlinear Spectroscopy (LENS), Via Nello Carrara 1, 50019 Sesto F.no (FI), Italy}
\affiliation{Department of Physics, Graduate School of Science, Kyoto University, Kyoto 606-8502, Japan}

%\author{Y.~Haruna}
%\affiliation{Department of Physics, Graduate School of Science, Kyoto University, Kyoto 606-8502, Japan}

\author{Y.~Takahashi}
\affiliation{Department of Physics, Graduate School of Science, Kyoto University, Kyoto 606-8502, Japan}

\date{\today}

\begin{abstract}
	We report on the experimental observation of Feshbach resonances in large
	mass-imbalance mixtures of Erbium (Er) and Lithium (Li). All combinations
	between \bEr, \bbEr\ and \bLi, \fLi\ are cooled to temperatures of a few
	microkelvin, partially by means of sympathetic cooling together with
	Ytterbium (Yb) as a third mixture component. The Er-Li inelastic
	interspecies collisional properties are studied for magnetic fields up to
	$680~\G$. In all cases resonant interspecies loss features, indicative of
	Feshbach resonances, have been observed. While most resonances have
	sub-Gauss widths, a few of them are broad and feature widths of several
	Gauss. Those broad resonances are a key to the realization of ultracold
	Er-Li quantum gas mixtures with tunable interactions.
\end{abstract}

\maketitle

\section{Introduction}
\label{sec:intro}

Ultracold quantum gases are by now an established and indispensable means to
manipulate, probe and study the intricate behavior of quantum matter as for
example impressively demonstrated in the realization of strongly correlated
quantum many-body systems~\cite{bloch_many-body_2008} as well as few-body
systems~\cite{naidon_efimov_2017}. In many cases experiments are facilitated
by use of Feshbach resonances that give rise to the necessary fine-tuned
control of atomic interactions~\cite{chin_feshbach_2010}. Landmark experiments
demonstrating the crossover from Bardeen-Cooper-Schrieffer (BCS) pairing to a
Bose-Einstein condensate (BEC)~\cite{randeria_crossover_2014} and Efimov
trimer-state series~\cite{zaccanti_observation_2009} are due to that minute
interaction control mechanism. Finally, going from single-component to
dual-species experiments the realm of accessible physics is further broadened
and allows as a direct application the simulation of impurity
problems~\cite{anderson_absence_1958, kondo_resistance_1964} and the study of
polaron physics~\cite{kohstall_metastability_2012} but also more intricate
applications such as the production of ultracold heteronuclear
molecules~\cite{chin_feshbach_2010}, mixed-species Efimov trimer physics with
reduced scaling constants~\cite{naidon_efimov_2017} and chiral p-wave
superfluids~\cite{nishida_induced_2009, wu_topological_2012}. It is important
to note that in the preceding last two examples the effects, while rendered
possible already by the presence of two different species, can be crucially
enhanced by use of large mass-imbalance mixtures. There, a large mass ratio is
the driving force in the reduction of Efimov scaling constants and a deciding
factor in bringing the superfluid critical temperature into experimentally
accessible realms.

We here present our first results of the experimental realization of a large
mass-imbalance mixture of heavy Erbium (Er) and light Lithium (Li) atoms and
the discovery of interspecies Feshbach resonances. In previous experiments
much effort was spent on similar investigations in ultracold mixtures of heavy
Ytterbium (Yb) and light Li~\cite{dowd_magnetic_2015,
schafer_spectroscopic_2017, green_feshbach_2020}. While these works scanned a
large range of isotope combinations, electronic configurations and magnetic
field strengths, only a few traces of Feshbach resonances could be detected
with limited applicability towards an efficient control of interspecies
interactions. In that respect Er seems a formidable candidate to remedy these
shortcomings of Yb-Li mixtures. While generally being comparable in mass to
the heavy lanthanide Yb, it has the distinct properties of a very large
magnetic moment giving rise to strong dipole-dipole interactions and the
existence of non-zero orbital angular momentum states. As such successful
efforts to create quantum degenerate gases of Er already reach ten years
back~\cite{aikawa_bose-einstein_2012} and lead to the unveiling of dense Er-Er
Feshbach spectra akin to chaotic behavior~\cite{frisch_quantum_2014}. Further
theoretical work revealed the strong possibility of a rich Feshbach physics in
Er-Li mixtures~\cite{gonzalez-martinez_magnetically_2015} which is the subject
of the present experimental work. Starting with the bosonic isotopes of either
\bEr\ or \bbEr\ and either bosonic \bLi\ or fermionic \fLi\ we are now able to
indeed identify a variety of resonantly enhanced inelastic scattering losses
between Er and Li at microkelvin temperatures. The purpose of the present work
is, after summarizing the experimental method in Sec.~\ref{sec:experiment} and
apart from generally reporting on the observation of Er-Li interspecies
Feshbach resonances, to give a succinct overview of the accessible parameters
and the range of observable resonances (Sec.~\ref{sec:results}). We conclude
in Sec.~\ref{sec:discussion} by discussing the range of possible applications
of such a large mass-imbalance mixture with tunable interspecies interactions.

\section{Experiment}
\label{sec:experiment}

The experiment is based on our Yb-Li machine described
previously~\cite{hara_quantum_2011, schafer_spin_2017}. In an upgrade of the
experiment a high-temperature oven for Er has additionally been added to the
setup, giving us the freedom to work with either single species or with
arbitrary combinations of Er, Yb and Li in dual- or triple-species mixtures.
The Er part of the setup and experimental sequence largely follows the
established techniques~\cite{aikawa_bose-einstein_2012}. From the Er source
heated to about $1050^\circ{\rm C}$ a hot atomic beam is formed. A transversal
cooling stage operating on the broad $4f^{12}6s^2\,(^3{\rm H}_6)$ to
$4f^{12}(^3{\rm H}_6)6s6p\,(^1{\rm P_1})$ transition at $401~\nm$ limits the
initial transverse spread to the thermal Er beam. A subsequent Zeeman slowing
stage operating on the same transition reduces the longitudinal velocity of
the Er atoms. A magneto-optical trap (MOT) operating on the narrow
$4f^{12}6s^2\,(^3{\rm H}_6)$ to $4f^{12}(^3{\rm H}_6)6s6p\,(^3{\rm P_1})$
intercombination line at $583~\nm$ forms the initial trapping stage. From
there the atoms are transferred into a far-off-resonant optical trap (FORT) of
crossed, focused beams at $1064~\nm$ and $1070~\nm$. It is known that due to
the narrow linewidth of the transition at $583~\nm$ of $186~\kHz$ and the
resulting displacement of the heavy Er atoms from the center of the MOT by
gravitational pull the atomic state is automatically polarized in the lowest
angular momentum sublevel~\cite{aikawa_bose-einstein_2012}, $m_J = -6$ in the
case at hand. By maintaining a sufficiently strong magnetic field for a well
defined quantization axis during the following forced evaporation period this
polarized state is maintained throughout the experiment. Evaporation typically
starts at $1.2~\G$ to prevent depolarization by thermal excitation and the
field is lowered to $0.4~\G$ once the sample is sufficiently cold for more
favorable collisional properties and better evaporation efficiency. Following
the technique outlined in~\cite{leroux_non-abelian_2018} the polarization
state has been confirmed via absorption imaging on the narrow $583~\nm$ line
which at moderate magnetic bias fields of about $20~\G$ and low imaging light
intensities is selective to a single spin state only. Trapping and cooling of
either \bLi\ or \fLi\ is performed as previously reported
\cite{schafer_experimental_2018, hara_quantum_2011} including optical pumping
of Li during the initial stage of the evaporation process to a spin-stretched
ground state. Successful spin polarization is confirmed by a Stern-Gerlach
technique, that is standard absorption imaging after short application of a
magnetic field gradient to spatially separate the $m_F$ hyperfine magnetic
sublevels. The accessible stretched states are $F = 1/2, m_F = \pm 1/2$ for
\fLi\ and $F = 1, m_F = \pm 1$ for \bLi. Additionally, when investigating
\bEr-\bLi\ a third loading stage of \bYb\ has been added to the experiment.
Due to the very efficient evaporation of \bYb\ and by virtue of sympathetic
cooling of both \bEr\ and \bLi\ larger and colder mixtures can be achieved.
The remaining Yb atoms are then removed before the actual Feshbach resonance
measurement by a short pulse of light resonant to the $^1{\rm S}_0$-$^1{\rm
P}_1$ transition of \bYb.

Forced evaporation of the two- or three-species mixture is done by suitably
lowering the intensities of the FORT lasers within $8~\s$ ($10~\s$ when
sympathetically cooled by Yb). After that time we typically obtain an Er-Li
mixture with $2\times10^4$ to $5\times10^4$ Er atoms and $5\times10^3$ to
$15\times10^3$ Li atoms. The actual atom number depends on the chosen isotope
combination, the desired final temperatures and also on the day-to-day
performance details of the experiment. The sample is non-degenerate and in the
absence of sympathetic cooling with Yb its temperature is found between $2$
and $5~\uK$ with Er typically being about $1~\uK$ colder than Li. This
demonstrates both that Er sympathetically cools down Li and that during the
last stages of the evaporation the transfer of thermal energy between the
species becomes less efficient which is similar to former experience in Yb-Li
mixtures~\cite{hansen_quantum_2011}, whereas in Yb-assisted sympathetic
cooling of \bEr-\bLi\ temperatures as low as $0.5~\uK$ are observed. The
expected trap frequencies are about $(\omega_x, \omega_y, \omega_z) = 2\pi
\times (420, 100, 50)~\Hz$ for Er and $2\pi \times (2640, 610, 300)~\Hz$ for
Li with the $z$-axis being in the vertical direction. In that condition the
difference in the trap centers due to gravitational sag is about $1.4~\micron$
which is sufficiently less than typical atom cloud dimensions in vertical
direction of about $6~\micron$ and is thus negligible for the purposes of the
present work.

The main part of the experimental sequence is to detect
magnetic-field-dependent changes in the interspecies inelastic collision rates
in a trap-loss sequence. For this, after the initial preparation of the
desired cold mixture sample a homogeneous magnetic field is linearly ramped up
to the target value within $10~\ms$. The system is then kept in this condition
for $500~\ms$. During this time inelastic collisions between the atoms lead to
a reduction of the number of atoms in the optical trap. The magnetic field
value is stabilized via a feedback signal from a current sensor on the
magnetic coil that acts on an insulated-gate bipolar transistor (IGBT). The
absolute magnetic field value is found by characterizing the narrow s-wave
Feshbach resonance~\cite{schunck_feshbach_2005} of \fLi\ at $543.28(8)~\G$ and
we assume linearity of the setup for other field values. Comparing our data
for \bEr-\bEr\ Feshbach resonances at low magnetic fields to those presented
in~\cite{frisch_quantum_2014} we find agreement in the observed resonance
positions to within $0.3~\G$. This systematic uncertainty is the major
contribution to our measurement error and outweighs the high frequency noise
of our magnetic field which is estimated to be about $0.02~\G$. The magnetic
field resolution is limited by the setpoint resolution of the employed 16~bit
digital-to-analog control frontend to about $0.03~\G$. After the holding time
at the target magnetic field the current in the magnetic coils is ramped down
to zero in $10~\ms$ and the system is kept there for another $10~\ms$ to allow
for eventual residual fields due to Eddy-currents to decay. In standard
time-of-flight absorption imaging we then image the remaining Er atoms after
$2~\ms$ and the Li atoms after $0.3~\ms$ of free expansion. Additionally, for
every magnetic field setting of this main sequence we interleave additional
control measurements in which either of the two species is removed from the
optical trap by a short pulse of resonant $583~\nm$ (Er) or $671~\nm$ (Li)
light before starting the magnetic field sequence. These measurements serve as
reference points to distinguish losses due to inter-species collisions from
those caused by single-species only interactions. Finally, while these data
are obtained every $60$ or $180~\mG$, typically every $1~\G$ additional data
of the mixture is taken where the magnetic field is kept constantly low. This
is to better detect possible drifts of the overall atom numbers or general
changes in the performance of the experimental setup.

\begin{figure*}[tb]
	\centering
	\includegraphics{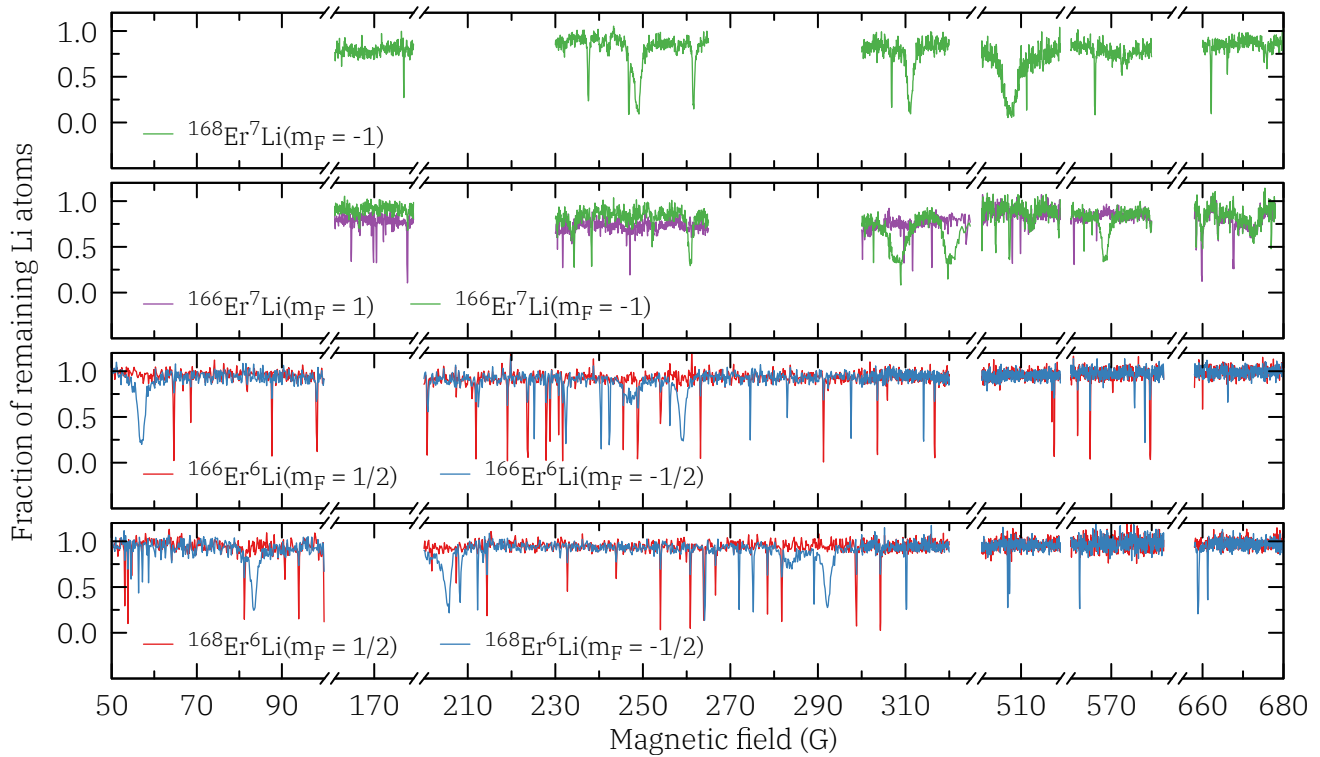}
	\caption{Magnetic field dependence of Li atom losses for various Er-Li
		mixtures. The four different panels show the fraction of remaining Li
		atoms for four Er-Li isotope combinations after $500~\ms$ interaction time
		at magnetic fields between $50$ and $680~\G$. (Note that the magnetic
		field axis is not continuous as some magnetic field ranges without data
		are omitted.) In each case Li has been optically pumped to one of the two
		stretched state configurations (see panel labels) and Er is always spin
		polarized into the $m_J = -6$ ground state. Each data point is the
		averaged ratio of typically three measurements with and without Er. Many
		narrow and some broad resonant loss features are observed.}
	\label{fig:fig1}
\end{figure*}

\section{Results}
\label{sec:results}

In the following we give an overview of the data obtained in the experiment.
This is to demonstrate the range of typically observed resonance shapes, their
densities and dependence on isotope- and spin-combinations.

We first take a look at the general isotope and hyperfine spin-state
dependence of the Feshbach spectra. Figure~\ref{fig:fig1} shows for the
magnetic field range $50$ to $680~\G$ trap-loss spectra of all isotope and
spin combinations that could be investigated in the present work. The spectra
are taken at a magnetic field resolution between $60$ and $180~\mG$. In each
case we report the fractional loss of Li atoms expressed as the ratio of atoms
remaining in the trap after the interaction time for the Er-Li mixture case
compared to the case of Li atoms only in the trap. Note that for brevity we
will in the following suppress $m_J = -6$ in the quantum state description of
Er and $F=1/2$ ($F=1$) for \fLi\ (\bLi), they are to be understood implicitly.
We further need to point out that for \bEr\bLi\ (top panel) we were not able
to create a cold mixture of \bEr\bLi$(m_F = 1)$. Even though at the beginning
of the evaporation ramp such a mixture could be produced, the \bLi$(m_F = 1)$
atoms are rapidly lost during the evaporation sequence. Also no magnetic bias
field could be found that allows for both efficient evaporation and
collisional stability of the mixture, indicating the existence of a broad
resonance at only a few Gauss for this isotope and hyperfine spin-state
combination. This is especially interesting as the energetically lowest state
in \bLi\ is the $m_F = 1$ state and naively one would expect this state to be
most stable towards possible inelastic processes. Instead it is the
energetically higher $m_F = -1$ state that can be prepared in the experiment.
The data in the top panel of Fig.~\ref{fig:fig1} therefore only shows the
\bEr\bLi$(m_F = -1)$ mixture. The other isotope combinations (remaining
panels) do not suffer from such a limitation. The available data do not cover
the complete magnetic field range (note the omission marks in the magnetic
field scale in Fig.~\ref{fig:fig1}) and only a total range of between roughly
$150~\G$ (\bLi) and $250~\G$ (\fLi) was scanned for each mixture. As such,
with a continuous measurement time of about a week per isotope and hyperfine
spin combination, a good overview of the spectral properties has been
obtained. Each mixture features a distinct resonance structure of numerous
narrow and several broader resonances. A detailed list of all observed
resonance positions and their widths is provided in
Tab.~\ref{tab:resonancelist}.
\begingroup
\squeezetable
\begin{table*}[ht]
	\caption{List of identified Er-Li interspecies Feshbach resonances. The
	resonance positions $B_0$ and their widths $\Delta B$ are given as
	determined by fits of squared Breit-Wigner lineshapes to the Li loss data
	shown in Fig.~\ref{fig:fig1}. Within the present experiments no reliable
	width determination is possible for very narrow resonances and those
	resonances with widths below $100~\mG$ are here indicated by $< 100$.
	Further, only resonances that are either reasonably broad or that cause a
	loss of Li atoms by at least one third are listed. Where imperfect optical
	pumping causes seemingly simultaneous losses in both Li spin states the
	resonance is attributed to the dominant spin component only. Resonances that
	are marginal for the identification are also included but put in brackets.
	Note that for $B > 50~\G$ only resonances within the investigated magnetic
	field ranges (cf.\ Fig.~\ref{fig:fig1}) are listed and that due to the
	finite magnetic field step size some very narrow resonances might not have
	been detected. Below $50~\G$ some additional resonances are included that
have been found in separate measurements.}
	\begin{tabular}{c}
		\hline
		\hline
\begin{minipage}[t]{0.23\textwidth}
\begin{tabular}[t]{c@{\extracolsep{5mm}}rr@{\extracolsep{5mm}}c}
& $B_0$ ($\G$) & $\Delta B$ ($\mG$) &\\
\hline
\multicolumn{4}{c}{\bf\boldmath${}^{168}$Er${}^{7}$Li($m_F = -1$\unboldmath)} \\
\hline
& $1.7$ & $<100$ &\\
& $7.5$ & $<100$ &\\
& $15.8$ & $100$ &\\
& $177.5$ & $<100$ &\\
& $237.6$ & $300$ &\\
& $246.8$ & $300$ &\\
& $248.8$ & $1600$ &\\
& $261.6$ & $500$ &\\
& $306.8$ & $200$ &\\
& $311.1$ & $1400$ &\\
& $507.3$ & $4000$ &\\
& $511.4$ & $<100$ &\\
& $566.0$ & $200$ &\\
& $662.1$ & $200$ &\\
& $666.1$ & $200$ &\\
& $675.9$ & $300$ &\\
\hline
\multicolumn{4}{c}{\bf\boldmath${}^{166}$Er${}^{7}$Li($m_F = 1$\unboldmath)} \\
\hline
& $13.1$ & $400$ &\\
& $16.8$ & $<100$ &\\
& $164.2$ & $100$ &\\
& $169.8$ & $100$ &\\
& $170.6$ & $100$ &\\
& $178.4$ & $200$ &\\
& $231.7$ & $<100$ &\\
& $233.6$ & $700$ &\\
& $247.0$ & $<100$ &\\
& $300.2$ & $100$ &\\
& $309.6$ & $<100$ &\\
& $311.6$ & $<100$ &\\
& $316.0$ & $100$ &\\
& $507.8$ & $100$ &\\
& $509.9$ & $<100$ &\\
& ($512.4$) & ($2300$) &\\
& $560.8$ & $100$ &\\
& $566.4$ & $<100$ &\\
& $658.5$ & $<100$ &\\
& $659.8$ & $300$ &\\
& $663.7$ & $300$ &\\
& $667.6$ & $300$ &\\
& ($672.4$) & ($3200$) &\\
\end{tabular}
\end{minipage}
\hfill
\begin{minipage}[t]{0.23\textwidth}
\begin{tabular}[t]{c@{\extracolsep{5mm}}rr@{\extracolsep{5mm}}c}
& $B_0$ ($\G$) & $\Delta B$ ($\mG$) &\\
\hline
\multicolumn{4}{c}{\bf\boldmath${}^{166}$Er${}^{7}$Li($m_F = -1$\unboldmath)} \\
\hline
& $23.6$ & $130$ &\\
& $233.8$ & $500$ &\\
& $234.3$ & $200$ &\\
& $238.3$ & $200$ &\\
& $260.8$ & $900$ &\\
& $300.7$ & $<100$ &\\
& $302.7$ & $<100$ &\\
& $306.3$ & $<100$ &\\
& $308.5$ & $4900$ &\\
& $308.9$ & $100$ &\\
& $319.8$ & $100$ &\\
& $319.9$ & $3200$ &\\
& $500.1$ & $<100$ &\\
& $503.6$ & $300$ &\\
& $507.1$ & $300$ &\\
& ($512.4$) & ($1700$) &\\
& $519.5$ & $400$ &\\
& $563.3$ & $<100$ &\\
& $565.3$ & $<100$ &\\
& $568.4$ & $1800$ &\\
& $658.4$ & $100$ &\\
& $663.7$ & $500$ &\\
& ($672.4$) & ($4000$) &\\
& $677.0$ & $<100$ &\\
\hline
\multicolumn{4}{c}{\bf\boldmath${}^{166}$Er${}^{6}$Li($m_F = 1/2$\unboldmath)} \\
\hline
& $18.9$ & $<100$ &\\
& $22.1$ & $<100$ &\\
& $25.6$ & $<100$ &\\
& $26.3$ & $<100$ &\\
& $37.8$ & $<100$ &\\
& $64.6$ & $200$ &\\
& $68.6$ & $<100$ &\\
& $87.7$ & $<100$ &\\
& $98.2$ & $<100$ &\\
& $200.8$ & $<100$ &\\
& $211.9$ & $100$ &\\
& $219.1$ & $100$ &\\
& $223.7$ & $<100$ &\\
& $228.0$ & $<100$ &\\
& $228.8$ & $100$ &\\
& $230.8$ & $<100$ &\\
\end{tabular}
\end{minipage}
\hfill
\begin{minipage}[t]{0.23\textwidth}
\begin{tabular}[t]{c@{\extracolsep{5mm}}rr@{\extracolsep{5mm}}c}
& $B_0$ ($\G$) & $\Delta B$ ($\mG$) &\\
\hline
& $231.7$ & $100$ &\\
& $245.5$ & $<100$ &\\
& $248.9$ & $<100$ &\\
& $254.1$ & $<100$ &\\
& $263.2$ & $<100$ &\\
& $291.3$ & $<100$ &\\
& $303.6$ & $<100$ &\\
& ($305.9$) & ($<100$) &\\
& $316.7$ & $200$ &\\
& $517.9$ & $<100$ &\\
& $518.4$ & $100$ &\\
& $561.7$ & $<100$ &\\
& $564.8$ & $100$ &\\
& ($570.4$) & ($100$) &\\
& $579.6$ & $200$ &\\
& $660.0$ & $<100$ &\\
\hline
\multicolumn{4}{c}{\bf\boldmath${}^{166}$Er${}^{6}$Li($m_F = -1/2$\unboldmath)} \\
\hline
& $13.3$ & $<100$ &\\
& $20.9$ & $<100$ &\\
& $24.0$ & $<100$ &\\
& $26.7$ & $100$ &\\
& $32.7$ & $<100$ &\\
& $57.0$ & $1700$ &\\
& $212.5$ & $300$ &\\
& $225.2$ & $<100$ &\\
& $232.4$ & $400$ &\\
& $240.4$ & $<100$ &\\
& $242.4$ & $<100$ &\\
& $247.0$ & $4200$ &\\
& $256.2$ & $<100$ &\\
& $259.0$ & $1300$ &\\
& $274.5$ & $<100$ &\\
& $283.0$ & $200$ &\\
& $297.5$ & $<100$ &\\
& $314.1$ & $100$ &\\
& $575.8$ & $200$ &\\
& $578.3$ & $100$ &\\
& $666.2$ & $<100$ &\\
\hline
\multicolumn{4}{c}{\bf\boldmath${}^{168}$Er${}^{6}$Li($m_F = 1/2$\unboldmath)} \\
\hline
& $23.2$ & $<100$ &\\
& ($42.5$) & ($<100$) &\\
& $46.4$ & $100$ &\\
& $53.1$ & $<100$ &\\
\end{tabular}
\end{minipage}
\hfill
\begin{minipage}[t]{0.23\textwidth}
\begin{tabular}[t]{c@{\extracolsep{5mm}}rr@{\extracolsep{5mm}}c}
& $B_0$ ($\G$) & $\Delta B$ ($\mG$) &\\
\hline
& $53.8$ & $<100$ &\\
& $54.7$ & $<100$ &\\
& $81.2$ & $200$ &\\
& $90.7$ & $<100$ &\\
& $93.9$ & $200$ &\\
& $201.8$ & $<100$ &\\
& $207.4$ & $<100$ &\\
& $214.4$ & $<100$ &\\
& $232.8$ & $100$ &\\
& $243.9$ & $100$ &\\
& $254.0$ & $<100$ &\\
& $260.9$ & $<100$ &\\
& $264.0$ & $<100$ &\\
& $266.6$ & $100$ &\\
& $278.5$ & $<100$ &\\
& $281.8$ & $<100$ &\\
& $298.7$ & $<100$ &\\
& $304.3$ & $200$ &\\
\hline
\multicolumn{4}{c}{\bf\boldmath${}^{168}$Er${}^{6}$Li($m_F = -1/2$\unboldmath)} \\
\hline
& $31.2$ & $100$ &\\
& $35.4$ & $<100$ &\\
& $41.5$ & $<100$ &\\
& ($42.5$) & ($<100$) &\\
& $54.6$ & $<100$ &\\
& $56.3$ & $<100$ &\\
& $57.2$ & $<100$ &\\
& $58.7$ & $<100$ &\\
& $83.4$ & $1500$ &\\
& $205.5$ & $1800$ &\\
& $208.1$ & $300$ &\\
& $212.3$ & $<100$ &\\
& $264.2$ & $<100$ &\\
& $272.0$ & $<100$ &\\
& $275.2$ & $200$ &\\
& $283.1$ & $4300$ &\\
& $289.1$ & $<100$ &\\
& $292.2$ & $1500$ &\\
& $310.2$ & $200$ &\\
& $506.6$ & $<100$ &\\
& $507.1$ & $100$ &\\
& $562.2$ & $100$ &\\
& $658.9$ & $300$ &\\
& $661.3$ & $100$ &\\
\end{tabular}
\end{minipage}\\
		\hline
		\hline
	\end{tabular}
	\label{tab:resonancelist}
\end{table*}
\endgroup
While the broader Feshbach resonances can be induced by ordinary s-wave
coupling, the numerously observed narrow resonances may be understood as a
result of anisotropy-induced Feshbach resonances where the anisotropy
originates form the electrostatic interaction between Er and
Li~\cite{kotochigova_controlling_2014, gonzalez-martinez_magnetically_2015}.
It seems further worth mentioning that there is an apparent lack of broad
Feshbach resonances with Li in the energetically lowest magnetic sublevel
(\bLi$(m_F = +1)$ and \fLi$(m_F = +1/2)$) which in not reflected in current
calculations~\cite{gonzalez-martinez_magnetically_2015}. Further experimental
efforts will be necessary, however, to further corroborate such observations.

\begin{figure}[b!]
	\centering
	\includegraphics{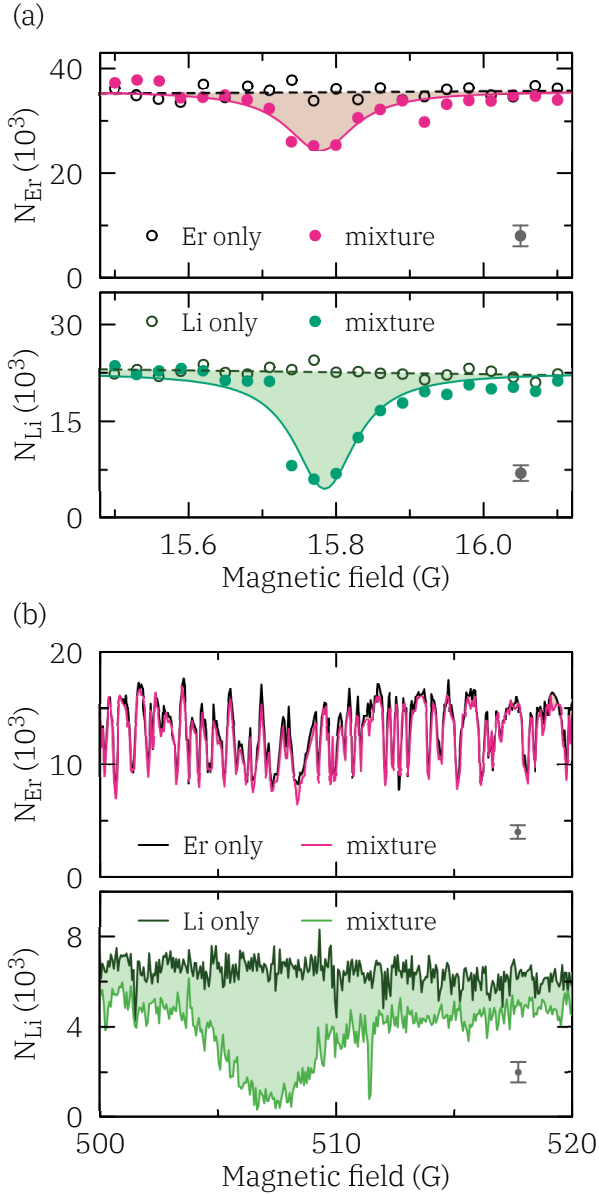}
	\caption{Detailed view of two \bEr\bLi$(m_F = -1)$ interspecies Feshbach
		resonances. The data acquisition and organization of each panel are as in
		Fig.~\ref{fig:fig1}. The typical statistical error of the displayed mean
		data is indicated in each panel by a single error bar example (dark gray).
		(a) At about $15.8~\G$ a narrow Feshbach resonance is observed. Resonant
		losses of Er and Li atoms are observed only in the mixture case, single
		species data show no loss indications. The solid (dashed) line indicates a
		Breit-Wigner (straight line) fit to the data and gives a FWHM width of
		about $0.1~\G$. The interaction time at each magnetic field was, different
		from the remaining experiments, set to only $10~\ms$. (b) At about
		$507~\G$ a broad interspecies Feshbach resonance with a width of about
		$4~\G$ is found and the resonant losses in \bLi\ are dominantly visible.
		The additional losses in the complicated structure of \bEr\ are only
		barely observable. At $511~\G$ an additional, very narrow interspecies
		loss feature is resolved.}
	\label{fig:fig2}
\end{figure}

We now turn to a discussion of the structures of individual resonances. The
panels of Fig.~\ref{fig:fig2} display in the mixture of \bEr\bLi$(m_F = -1)$ a
narrow Feshbach resonance at low magnetic fields around $15.8~\G$ and a
particularly broad resonance at higher fields around $510~\G$. In both cases
the mixture sample is prepared at a temperature of about $2.0~\uK$ for \bEr\
and $2.7~\uK$ for \bLi. In the vicinity of $15.8~\G$ Fig.~\ref{fig:fig2}(a)
shows a narrow interspecies Feshbach resonance with a full width at half
maximum (FWHM) of about $0.1~\G$. In the shown magnetic field range the single
species data (open symbols), where the atoms of the other species were removed
from the trap before setting the magnetic field, is featureless. In the
mixture, however, a coinciding loss of both Er and Li is visible, indicative
of the enhanced inelastic losses of an interspecies Feshbach resonance. As the
interaction time for that particular experiment has been set to only $10~\ms$,
the inelastic losses seem unusually large for this particular resonance. Note
that the remaining experiments reported in the present work operate at a
holding time of $500~\ms$. In the higher magnetic field range of
Fig.~\ref{fig:fig2}(b), looking first at the \bEr\ data, the dominant and
previously reported dense structure of Er-Er Feshbach resonances is
evident~\cite{frisch_quantum_2014}. Going to the \bLi\ data again the
resonance structure relevant to the present research is revealed: While the
``Li only'' data shows only a smooth variation in the overall atom number that
can be fully attributed to a drift of the general performance of the
experiment, the data of the mixture clearly display a broad resonant loss. In
contrast to the data at $15.8~\G$, the interspecies loss is this time only
weakly reflected in the Er data. This is partially due to a reduced visibility
caused by the dense Er-Er Feshbach resonance structure but is expected to also be
induced by the large mass imbalance between the two species. The resonance
shape is asymmetric and we extract an approximate resonance position and FWHM
by a Lorentzian fit to the \bLi\ data. We find a resonance position of
$507.3~\G$ and a FWHM of $4.0~\G$. This broad resonance is accompanied by a
very narrow one at $511.4~\G$ that has a width well below $0.1~\G$ at the
limit of our experimental resolution.

\section{Discussion and Conclusion}
\label{sec:discussion}

With the present set of experimental results we demonstrate the realization of
large mass-imbalance mixtures of Er and Li in various isotopic combinations.
While not seamlessly exploring the complete magnetic field range up to
$680~\G$ we focus on a more detailed investigation of the Er-Li interspecies
Feshbach resonance structure in multiple field locations. This approach is
chosen to obtain a general idea of the expected spectra and to better support
theoretical efforts in the modeling and understanding of the Feshbach spectra
by providing data samples over a possibly large range of magnetic fields,
isotope and spin combinations. As expected we find a wealth of interspecies
Feshbach resonances at a density of about one resonance per ten Gauss, most of
them quite narrow, some of them with widths beyond $1~\G$. This encouraging
overall consistency between the observed Feshbach resonance spectra and
current theory predictions~\cite{gonzalez-martinez_magnetically_2015} is
contrasted by the lack of a detailed understanding of
the exact resonance structures. Joint efforts of both experiment and theory
will be necessary for a better grasp of this complicated system. Additionally,
the present work reveals mixture lifetimes on the order of up to $1~\s$ even
at resonance. This is, e.g., about two orders of magnitude longer than
inelastic collision rate limited lifetimes found in the large mass-imbalance
system of Yb(${}^3\mathrm{P}_2)$-Li~\cite{schafer_spectroscopic_2017} and an
encouraging starting point for further research of Er-Li mixtures in the
strongly interacting regime.

To our knowledge this mixture system currently represents the largest
experimentally realized mass-imbalanced mixture with tunable interactions.
Additional measurements are in preparation to precisely determine the binding
energies of the corresponding Feshbach molecules. Considering in particular a
similar kind of mixture of heavy \fEr\ fermions and light \bLi\ bosons this
opens the road to three-body Efimov states with both reduced scaling constants
and non-vanishing angular momenta~\cite{naidon_efimov_2017}, and chiral p-wave
superfluids in mixed dimensions. Additionally, ultracold mixtures of Yb and Er
should allow for a detailed study of mass-scaling effects in Yb-Er Feshbach
resonances~\cite{kosicki_quantum_2020}. Ultracold triple-species quantum gas
mixtures of Ytterbium, Erbium and Lithium are also within reach.

\section*{Acknowledgments}
We thank Y.\ Takasu for early experimental assistance and useful discussions,
K.\ Aikawa for helpful advise in early stages of the experiment and P.\
Żuchowski for fruitful discussions of the Feshbach spectra and for sharing his
theoretical insight with us. NM acknowledges support from the JSPS (KAKENHI
Grant No.\ 21J20153). This work was supported by the Grant-in-Aid for
Scientific Research of JSPS Grants No.\ JP17H06138, No.\ 18H05405, and No.\
18H05228, JST CREST Grant No.\ JPMJCR1673 and the Impulsing Paradigm Change
through Disruptive Technologies (ImPACT) program by the Cabinet Office,
Government of Japan, and MEXT Quantum Leap Flagship Program (MEXT Q-LEAP)
Grant No.\ JPMXS0118069021.

%\bibliographystyle{apsrev4-1}
%\bibliography{ErLiFeshbach}

\begin{thebibliography}{24}%
\makeatletter
\providecommand \@ifxundefined [1]{%
 \@ifx{#1\undefined}
}%
\providecommand \@ifnum [1]{%
 \ifnum #1\expandafter \@firstoftwo
 \else \expandafter \@secondoftwo
 \fi
}%
\providecommand \@ifx [1]{%
 \ifx #1\expandafter \@firstoftwo
 \else \expandafter \@secondoftwo
 \fi
}%
\providecommand \natexlab [1]{#1}%
\providecommand \enquote  [1]{``#1''}%
\providecommand \bibnamefont  [1]{#1}%
\providecommand \bibfnamefont [1]{#1}%
\providecommand \citenamefont [1]{#1}%
\providecommand \href@noop [0]{\@secondoftwo}%
\providecommand \href [0]{\begingroup \@sanitize@url \@href}%
\providecommand \@href[1]{\@@startlink{#1}\@@href}%
\providecommand \@@href[1]{\endgroup#1\@@endlink}%
\providecommand \@sanitize@url [0]{\catcode `\\12\catcode `\$12\catcode
  `\&12\catcode `\#12\catcode `\^12\catcode `\_12\catcode `\%12\relax}%
\providecommand \@@startlink[1]{}%
\providecommand \@@endlink[0]{}%
\providecommand \url  [0]{\begingroup\@sanitize@url \@url }%
\providecommand \@url [1]{\endgroup\@href {#1}{\urlprefix }}%
\providecommand \urlprefix  [0]{URL }%
\providecommand \Eprint [0]{\href }%
\providecommand \doibase [0]{http://dx.doi.org/}%
\providecommand \selectlanguage [0]{\@gobble}%
\providecommand \bibinfo  [0]{\@secondoftwo}%
\providecommand \bibfield  [0]{\@secondoftwo}%
\providecommand \translation [1]{[#1]}%
\providecommand \BibitemOpen [0]{}%
\providecommand \bibitemStop [0]{}%
\providecommand \bibitemNoStop [0]{.\EOS\space}%
\providecommand \EOS [0]{\spacefactor3000\relax}%
\providecommand \BibitemShut  [1]{\csname bibitem#1\endcsname}%
\let\auto@bib@innerbib\@empty
%</preamble>
\bibitem [{\citenamefont {Bloch}\ \emph {et~al.}(2008)\citenamefont {Bloch},
  \citenamefont {Dalibard},\ and\ \citenamefont
  {Zwerger}}]{bloch_many-body_2008}%
  \BibitemOpen
  \bibfield  {author} {\bibinfo {author} {\bibfnamefont {I.}~\bibnamefont
  {Bloch}}, \bibinfo {author} {\bibfnamefont {J.}~\bibnamefont {Dalibard}}, \
  and\ \bibinfo {author} {\bibfnamefont {W.}~\bibnamefont {Zwerger}},\ }\href
  {\doibase 10.1103/RevModPhys.80.885} {\bibfield  {journal} {\bibinfo
  {journal} {Rev. Mod. Phys.}\ }\textbf {\bibinfo {volume} {80}},\ \bibinfo
  {pages} {885} (\bibinfo {year} {2008})}\BibitemShut {NoStop}%
\bibitem [{\citenamefont {Naidon}\ and\ \citenamefont
  {Endo}(2017)}]{naidon_efimov_2017}%
  \BibitemOpen
  \bibfield  {author} {\bibinfo {author} {\bibfnamefont {P.}~\bibnamefont
  {Naidon}}\ and\ \bibinfo {author} {\bibfnamefont {S.}~\bibnamefont {Endo}},\
  }\href {\doibase 10.1088/1361-6633/aa50e8} {\bibfield  {journal} {\bibinfo
  {journal} {Rep. Prog. Phys.}\ }\textbf {\bibinfo {volume} {80}},\ \bibinfo
  {pages} {056001} (\bibinfo {year} {2017})}\BibitemShut {NoStop}%
\bibitem [{\citenamefont {Chin}\ \emph {et~al.}(2010)\citenamefont {Chin},
  \citenamefont {Grimm}, \citenamefont {Julienne},\ and\ \citenamefont
  {Tiesinga}}]{chin_feshbach_2010}%
  \BibitemOpen
  \bibfield  {author} {\bibinfo {author} {\bibfnamefont {C.}~\bibnamefont
  {Chin}}, \bibinfo {author} {\bibfnamefont {R.}~\bibnamefont {Grimm}},
  \bibinfo {author} {\bibfnamefont {P.}~\bibnamefont {Julienne}}, \ and\
  \bibinfo {author} {\bibfnamefont {E.}~\bibnamefont {Tiesinga}},\ }\href
  {\doibase 10.1103/RevModPhys.82.1225} {\bibfield  {journal} {\bibinfo
  {journal} {Rev. Mod. Phys.}\ }\textbf {\bibinfo {volume} {82}},\ \bibinfo
  {pages} {1225} (\bibinfo {year} {2010})}\BibitemShut {NoStop}%
\bibitem [{\citenamefont {Randeria}\ and\ \citenamefont
  {Taylor}(2014)}]{randeria_crossover_2014}%
  \BibitemOpen
  \bibfield  {author} {\bibinfo {author} {\bibfnamefont {M.}~\bibnamefont
  {Randeria}}\ and\ \bibinfo {author} {\bibfnamefont {E.}~\bibnamefont
  {Taylor}},\ }\href {\doibase 10.1146/annurev-conmatphys-031113-133829}
  {\bibfield  {journal} {\bibinfo  {journal} {Annu. Rev. Condens. Matter
  Phys.}\ }\textbf {\bibinfo {volume} {5}},\ \bibinfo {pages} {209} (\bibinfo
  {year} {2014})}\BibitemShut {NoStop}%
\bibitem [{\citenamefont {Zaccanti}\ \emph {et~al.}(2009)\citenamefont
  {Zaccanti}, \citenamefont {Deissler}, \citenamefont
  {D{\textquoteright}Errico}, \citenamefont {Fattori}, \citenamefont
  {Jona-Lasinio}, \citenamefont {M{\"u}ller}, \citenamefont {Roati},
  \citenamefont {Inguscio},\ and\ \citenamefont
  {Modugno}}]{zaccanti_observation_2009}%
  \BibitemOpen
  \bibfield  {author} {\bibinfo {author} {\bibfnamefont {M.}~\bibnamefont
  {Zaccanti}}, \bibinfo {author} {\bibfnamefont {B.}~\bibnamefont {Deissler}},
  \bibinfo {author} {\bibfnamefont {C.}~\bibnamefont
  {D{\textquoteright}Errico}}, \bibinfo {author} {\bibfnamefont
  {M.}~\bibnamefont {Fattori}}, \bibinfo {author} {\bibfnamefont
  {M.}~\bibnamefont {Jona-Lasinio}}, \bibinfo {author} {\bibfnamefont
  {S.}~\bibnamefont {M{\"u}ller}}, \bibinfo {author} {\bibfnamefont
  {G.}~\bibnamefont {Roati}}, \bibinfo {author} {\bibfnamefont
  {M.}~\bibnamefont {Inguscio}}, \ and\ \bibinfo {author} {\bibfnamefont
  {G.}~\bibnamefont {Modugno}},\ }\href {\doibase 10.1038/nphys1334} {\bibfield
   {journal} {\bibinfo  {journal} {Nat. Phys.}\ }\textbf {\bibinfo {volume}
  {5}},\ \bibinfo {pages} {586} (\bibinfo {year} {2009})}\BibitemShut {NoStop}%
\bibitem [{\citenamefont {Anderson}(1958)}]{anderson_absence_1958}%
  \BibitemOpen
  \bibfield  {author} {\bibinfo {author} {\bibfnamefont {P.~W.}\ \bibnamefont
  {Anderson}},\ }\href {\doibase 10.1103/PhysRev.109.1492} {\bibfield
  {journal} {\bibinfo  {journal} {Phys. Rev.}\ }\textbf {\bibinfo {volume}
  {109}},\ \bibinfo {pages} {1492} (\bibinfo {year} {1958})}\BibitemShut
  {NoStop}%
\bibitem [{\citenamefont {Kondo}(1964)}]{kondo_resistance_1964}%
  \BibitemOpen
  \bibfield  {author} {\bibinfo {author} {\bibfnamefont {J.}~\bibnamefont
  {Kondo}},\ }\href {\doibase 10.1143/PTP.32.37} {\bibfield  {journal}
  {\bibinfo  {journal} {Prog. Theor. Phys.}\ }\textbf {\bibinfo {volume}
  {32}},\ \bibinfo {pages} {37} (\bibinfo {year} {1964})}\BibitemShut {NoStop}%
\bibitem [{\citenamefont {Kohstall}\ \emph {et~al.}(2012)\citenamefont
  {Kohstall}, \citenamefont {Zaccanti}, \citenamefont {Jag}, \citenamefont
  {Trenkwalder}, \citenamefont {Massignan}, \citenamefont {Bruun},
  \citenamefont {Schreck},\ and\ \citenamefont
  {Grimm}}]{kohstall_metastability_2012}%
  \BibitemOpen
  \bibfield  {author} {\bibinfo {author} {\bibfnamefont {C.}~\bibnamefont
  {Kohstall}}, \bibinfo {author} {\bibfnamefont {M.}~\bibnamefont {Zaccanti}},
  \bibinfo {author} {\bibfnamefont {M.}~\bibnamefont {Jag}}, \bibinfo {author}
  {\bibfnamefont {A.}~\bibnamefont {Trenkwalder}}, \bibinfo {author}
  {\bibfnamefont {P.}~\bibnamefont {Massignan}}, \bibinfo {author}
  {\bibfnamefont {G.~M.}\ \bibnamefont {Bruun}}, \bibinfo {author}
  {\bibfnamefont {F.}~\bibnamefont {Schreck}}, \ and\ \bibinfo {author}
  {\bibfnamefont {R.}~\bibnamefont {Grimm}},\ }\href {\doibase
  10.1038/nature11065} {\bibfield  {journal} {\bibinfo  {journal} {Nature}\
  }\textbf {\bibinfo {volume} {485}},\ \bibinfo {pages} {615} (\bibinfo {year}
  {2012})}\BibitemShut {NoStop}%
\bibitem [{\citenamefont {Nishida}(2009)}]{nishida_induced_2009}%
  \BibitemOpen
  \bibfield  {author} {\bibinfo {author} {\bibfnamefont {Y.}~\bibnamefont
  {Nishida}},\ }\href {\doibase 10.1016/j.aop.2008.10.011} {\bibfield
  {journal} {\bibinfo  {journal} {Ann. Phys.}\ }\textbf {\bibinfo {volume}
  {324}},\ \bibinfo {pages} {897} (\bibinfo {year} {2009})}\BibitemShut
  {NoStop}%
\bibitem [{\citenamefont {Wu}\ \emph {et~al.}(2012)\citenamefont {Wu},
  \citenamefont {He}, \citenamefont {Zang},\ and\ \citenamefont
  {Kou}}]{wu_topological_2012}%
  \BibitemOpen
  \bibfield  {author} {\bibinfo {author} {\bibfnamefont {Y.-J.}\ \bibnamefont
  {Wu}}, \bibinfo {author} {\bibfnamefont {J.}~\bibnamefont {He}}, \bibinfo
  {author} {\bibfnamefont {C.-L.}\ \bibnamefont {Zang}}, \ and\ \bibinfo
  {author} {\bibfnamefont {S.-P.}\ \bibnamefont {Kou}},\ }\href {\doibase
  10.1103/PhysRevB.86.085128} {\bibfield  {journal} {\bibinfo  {journal} {Phys.
  Rev. B}\ }\textbf {\bibinfo {volume} {86}},\ \bibinfo {pages} {085128}
  (\bibinfo {year} {2012})}\BibitemShut {NoStop}%
\bibitem [{\citenamefont {Dowd}\ \emph {et~al.}(2015)\citenamefont {Dowd},
  \citenamefont {Roy}, \citenamefont {Shrestha}, \citenamefont {Petrov},
  \citenamefont {Makrides}, \citenamefont {Kotochigova},\ and\ \citenamefont
  {Gupta}}]{dowd_magnetic_2015}%
  \BibitemOpen
  \bibfield  {author} {\bibinfo {author} {\bibfnamefont {W.}~\bibnamefont
  {Dowd}}, \bibinfo {author} {\bibfnamefont {R.~J.}\ \bibnamefont {Roy}},
  \bibinfo {author} {\bibfnamefont {R.~K.}\ \bibnamefont {Shrestha}}, \bibinfo
  {author} {\bibfnamefont {A.}~\bibnamefont {Petrov}}, \bibinfo {author}
  {\bibfnamefont {C.}~\bibnamefont {Makrides}}, \bibinfo {author}
  {\bibfnamefont {S.}~\bibnamefont {Kotochigova}}, \ and\ \bibinfo {author}
  {\bibfnamefont {S.}~\bibnamefont {Gupta}},\ }\href {\doibase
  10.1088/1367-2630/17/5/055007} {\bibfield  {journal} {\bibinfo  {journal}
  {New J. Phys.}\ }\textbf {\bibinfo {volume} {17}},\ \bibinfo {pages} {055007}
  (\bibinfo {year} {2015})}\BibitemShut {NoStop}%
\bibitem [{\citenamefont {Sch{\"a}fer}\ \emph
  {et~al.}(2017{\natexlab{a}})\citenamefont {Sch{\"a}fer}, \citenamefont
  {Konishi}, \citenamefont {Bouscal}, \citenamefont {Yagami},\ and\
  \citenamefont {Takahashi}}]{schafer_spectroscopic_2017}%
  \BibitemOpen
  \bibfield  {author} {\bibinfo {author} {\bibfnamefont {F.}~\bibnamefont
  {Sch{\"a}fer}}, \bibinfo {author} {\bibfnamefont {H.}~\bibnamefont
  {Konishi}}, \bibinfo {author} {\bibfnamefont {A.}~\bibnamefont {Bouscal}},
  \bibinfo {author} {\bibfnamefont {T.}~\bibnamefont {Yagami}}, \ and\ \bibinfo
  {author} {\bibfnamefont {Y.}~\bibnamefont {Takahashi}},\ }\href {\doibase
  10.1103/PhysRevA.96.032711} {\bibfield  {journal} {\bibinfo  {journal} {Phys.
  Rev. A}\ }\textbf {\bibinfo {volume} {96}},\ \bibinfo {pages} {032711}
  (\bibinfo {year} {2017}{\natexlab{a}})}\BibitemShut {NoStop}%
\bibitem [{\citenamefont {Green}\ \emph {et~al.}(2020)\citenamefont {Green},
  \citenamefont {Li}, \citenamefont {See~Toh}, \citenamefont {Tang},
  \citenamefont {McCormick}, \citenamefont {Li}, \citenamefont {Tiesinga},
  \citenamefont {Kotochigova},\ and\ \citenamefont
  {Gupta}}]{green_feshbach_2020}%
  \BibitemOpen
  \bibfield  {author} {\bibinfo {author} {\bibfnamefont {A.}~\bibnamefont
  {Green}}, \bibinfo {author} {\bibfnamefont {H.}~\bibnamefont {Li}}, \bibinfo
  {author} {\bibfnamefont {J.~H.}\ \bibnamefont {See~Toh}}, \bibinfo {author}
  {\bibfnamefont {X.}~\bibnamefont {Tang}}, \bibinfo {author} {\bibfnamefont
  {K.~C.}\ \bibnamefont {McCormick}}, \bibinfo {author} {\bibfnamefont
  {M.}~\bibnamefont {Li}}, \bibinfo {author} {\bibfnamefont {E.}~\bibnamefont
  {Tiesinga}}, \bibinfo {author} {\bibfnamefont {S.}~\bibnamefont
  {Kotochigova}}, \ and\ \bibinfo {author} {\bibfnamefont {S.}~\bibnamefont
  {Gupta}},\ }\href {\doibase 10.1103/PhysRevX.10.031037} {\bibfield  {journal}
  {\bibinfo  {journal} {Phys. Rev. X}\ }\textbf {\bibinfo {volume} {10}},\
  \bibinfo {pages} {031037} (\bibinfo {year} {2020})}\BibitemShut {NoStop}%
\bibitem [{\citenamefont {Aikawa}\ \emph {et~al.}(2012)\citenamefont {Aikawa},
  \citenamefont {Frisch}, \citenamefont {Mark}, \citenamefont {Baier},
  \citenamefont {Rietzler}, \citenamefont {Grimm},\ and\ \citenamefont
  {Ferlaino}}]{aikawa_bose-einstein_2012}%
  \BibitemOpen
  \bibfield  {author} {\bibinfo {author} {\bibfnamefont {K.}~\bibnamefont
  {Aikawa}}, \bibinfo {author} {\bibfnamefont {A.}~\bibnamefont {Frisch}},
  \bibinfo {author} {\bibfnamefont {M.}~\bibnamefont {Mark}}, \bibinfo {author}
  {\bibfnamefont {S.}~\bibnamefont {Baier}}, \bibinfo {author} {\bibfnamefont
  {A.}~\bibnamefont {Rietzler}}, \bibinfo {author} {\bibfnamefont
  {R.}~\bibnamefont {Grimm}}, \ and\ \bibinfo {author} {\bibfnamefont
  {F.}~\bibnamefont {Ferlaino}},\ }\href {\doibase
  10.1103/PhysRevLett.108.210401} {\bibfield  {journal} {\bibinfo  {journal}
  {Phys. Rev. Lett.}\ }\textbf {\bibinfo {volume} {108}},\ \bibinfo {pages}
  {210401} (\bibinfo {year} {2012})}\BibitemShut {NoStop}%
\bibitem [{\citenamefont {Frisch}\ \emph {et~al.}(2014)\citenamefont {Frisch},
  \citenamefont {Mark}, \citenamefont {Aikawa}, \citenamefont {Ferlaino},
  \citenamefont {Bohn}, \citenamefont {Makrides}, \citenamefont {Petrov},\ and\
  \citenamefont {Kotochigova}}]{frisch_quantum_2014}%
  \BibitemOpen
  \bibfield  {author} {\bibinfo {author} {\bibfnamefont {A.}~\bibnamefont
  {Frisch}}, \bibinfo {author} {\bibfnamefont {M.}~\bibnamefont {Mark}},
  \bibinfo {author} {\bibfnamefont {K.}~\bibnamefont {Aikawa}}, \bibinfo
  {author} {\bibfnamefont {F.}~\bibnamefont {Ferlaino}}, \bibinfo {author}
  {\bibfnamefont {J.~L.}\ \bibnamefont {Bohn}}, \bibinfo {author}
  {\bibfnamefont {C.}~\bibnamefont {Makrides}}, \bibinfo {author}
  {\bibfnamefont {A.}~\bibnamefont {Petrov}}, \ and\ \bibinfo {author}
  {\bibfnamefont {S.}~\bibnamefont {Kotochigova}},\ }\href {\doibase
  10.1038/nature13137} {\bibfield  {journal} {\bibinfo  {journal} {Nature}\
  }\textbf {\bibinfo {volume} {507}},\ \bibinfo {pages} {475} (\bibinfo {year}
  {2014})}\BibitemShut {NoStop}%
\bibitem [{\citenamefont {Gonz{\'a}lez-Mart{\'i}nez}\ and\ \citenamefont
  {{\.Z}uchowski}(2015)}]{gonzalez-martinez_magnetically_2015}%
  \BibitemOpen
  \bibfield  {author} {\bibinfo {author} {\bibfnamefont {M.~L.}\ \bibnamefont
  {Gonz{\'a}lez-Mart{\'i}nez}}\ and\ \bibinfo {author} {\bibfnamefont {P.~S.}\
  \bibnamefont {{\.Z}uchowski}},\ }\href {\doibase 10.1103/PhysRevA.92.022708}
  {\bibfield  {journal} {\bibinfo  {journal} {Phys. Rev. A}\ }\textbf {\bibinfo
  {volume} {92}},\ \bibinfo {pages} {022708} (\bibinfo {year}
  {2015})}\BibitemShut {NoStop}%
\bibitem [{\citenamefont {Hara}\ \emph {et~al.}(2011)\citenamefont {Hara},
  \citenamefont {Takasu}, \citenamefont {Yamaoka}, \citenamefont {Doyle},\ and\
  \citenamefont {Takahashi}}]{hara_quantum_2011}%
  \BibitemOpen
  \bibfield  {author} {\bibinfo {author} {\bibfnamefont {H.}~\bibnamefont
  {Hara}}, \bibinfo {author} {\bibfnamefont {Y.}~\bibnamefont {Takasu}},
  \bibinfo {author} {\bibfnamefont {Y.}~\bibnamefont {Yamaoka}}, \bibinfo
  {author} {\bibfnamefont {J.~M.}\ \bibnamefont {Doyle}}, \ and\ \bibinfo
  {author} {\bibfnamefont {Y.}~\bibnamefont {Takahashi}},\ }\href {\doibase
  10.1103/PhysRevLett.106.205304} {\bibfield  {journal} {\bibinfo  {journal}
  {Phys. Rev. Lett.}\ }\textbf {\bibinfo {volume} {106}},\ \bibinfo {pages}
  {205304} (\bibinfo {year} {2011})}\BibitemShut {NoStop}%
\bibitem [{\citenamefont {Sch{\"a}fer}\ \emph
  {et~al.}(2017{\natexlab{b}})\citenamefont {Sch{\"a}fer}, \citenamefont
  {Konishi}, \citenamefont {Bouscal}, \citenamefont {Yagami},\ and\
  \citenamefont {Takahashi}}]{schafer_spin_2017}%
  \BibitemOpen
  \bibfield  {author} {\bibinfo {author} {\bibfnamefont {F.}~\bibnamefont
  {Sch{\"a}fer}}, \bibinfo {author} {\bibfnamefont {H.}~\bibnamefont
  {Konishi}}, \bibinfo {author} {\bibfnamefont {A.}~\bibnamefont {Bouscal}},
  \bibinfo {author} {\bibfnamefont {T.}~\bibnamefont {Yagami}}, \ and\ \bibinfo
  {author} {\bibfnamefont {Y.}~\bibnamefont {Takahashi}},\ }\href {\doibase
  10.1088/1367-2630/aa8cec} {\bibfield  {journal} {\bibinfo  {journal} {New J.
  Phys.}\ }\textbf {\bibinfo {volume} {19}},\ \bibinfo {pages} {103039}
  (\bibinfo {year} {2017}{\natexlab{b}})}\BibitemShut {NoStop}%
\bibitem [{\citenamefont {Leroux}\ \emph {et~al.}(2018)\citenamefont {Leroux},
  \citenamefont {Pandey}, \citenamefont {Rehbi}, \citenamefont {Chevy},
  \citenamefont {Miniatura}, \citenamefont {Gr{\'e}maud},\ and\ \citenamefont
  {Wilkowski}}]{leroux_non-abelian_2018}%
  \BibitemOpen
  \bibfield  {author} {\bibinfo {author} {\bibfnamefont {F.}~\bibnamefont
  {Leroux}}, \bibinfo {author} {\bibfnamefont {K.}~\bibnamefont {Pandey}},
  \bibinfo {author} {\bibfnamefont {R.}~\bibnamefont {Rehbi}}, \bibinfo
  {author} {\bibfnamefont {F.}~\bibnamefont {Chevy}}, \bibinfo {author}
  {\bibfnamefont {C.}~\bibnamefont {Miniatura}}, \bibinfo {author}
  {\bibfnamefont {B.}~\bibnamefont {Gr{\'e}maud}}, \ and\ \bibinfo {author}
  {\bibfnamefont {D.}~\bibnamefont {Wilkowski}},\ }\href {\doibase
  10.1038/s41467-018-05865-3} {\bibfield  {journal} {\bibinfo  {journal} {Nat.
  Comm.}\ }\textbf {\bibinfo {volume} {9}},\ \bibinfo {pages} {3580} (\bibinfo
  {year} {2018})}\BibitemShut {NoStop}%
\bibitem [{\citenamefont {Sch{\"a}fer}\ \emph {et~al.}(2018)\citenamefont
  {Sch{\"a}fer}, \citenamefont {Mizukami}, \citenamefont {Yu}, \citenamefont
  {Koibuchi}, \citenamefont {Bouscal},\ and\ \citenamefont
  {Takahashi}}]{schafer_experimental_2018}%
  \BibitemOpen
  \bibfield  {author} {\bibinfo {author} {\bibfnamefont {F.}~\bibnamefont
  {Sch{\"a}fer}}, \bibinfo {author} {\bibfnamefont {N.}~\bibnamefont
  {Mizukami}}, \bibinfo {author} {\bibfnamefont {P.}~\bibnamefont {Yu}},
  \bibinfo {author} {\bibfnamefont {S.}~\bibnamefont {Koibuchi}}, \bibinfo
  {author} {\bibfnamefont {A.}~\bibnamefont {Bouscal}}, \ and\ \bibinfo
  {author} {\bibfnamefont {Y.}~\bibnamefont {Takahashi}},\ }\href {\doibase
  10.1103/PhysRevA.98.051602} {\bibfield  {journal} {\bibinfo  {journal} {Phys.
  Rev. A}\ }\textbf {\bibinfo {volume} {98}},\ \bibinfo {pages} {051602(R)}
  (\bibinfo {year} {2018})}\BibitemShut {NoStop}%
\bibitem [{\citenamefont {Hansen}\ \emph {et~al.}(2011)\citenamefont {Hansen},
  \citenamefont {Khramov}, \citenamefont {Dowd}, \citenamefont {Jamison},
  \citenamefont {Ivanov},\ and\ \citenamefont {Gupta}}]{hansen_quantum_2011}%
  \BibitemOpen
  \bibfield  {author} {\bibinfo {author} {\bibfnamefont {A.~H.}\ \bibnamefont
  {Hansen}}, \bibinfo {author} {\bibfnamefont {A.}~\bibnamefont {Khramov}},
  \bibinfo {author} {\bibfnamefont {W.~H.}\ \bibnamefont {Dowd}}, \bibinfo
  {author} {\bibfnamefont {A.~O.}\ \bibnamefont {Jamison}}, \bibinfo {author}
  {\bibfnamefont {V.~V.}\ \bibnamefont {Ivanov}}, \ and\ \bibinfo {author}
  {\bibfnamefont {S.}~\bibnamefont {Gupta}},\ }\href {\doibase
  10.1103/PhysRevA.84.011606} {\bibfield  {journal} {\bibinfo  {journal} {Phys.
	Rev. A}\ }\textbf {\bibinfo {volume} {84}},\ \bibinfo {pages} {011606(R)}
  (\bibinfo {year} {2011})}\BibitemShut {NoStop}%
\bibitem [{\citenamefont {Schunck}\ \emph {et~al.}(2005)\citenamefont
  {Schunck}, \citenamefont {Zwierlein}, \citenamefont {Stan}, \citenamefont
  {Raupach}, \citenamefont {Ketterle}, \citenamefont {Simoni}, \citenamefont
  {Tiesinga}, \citenamefont {Williams},\ and\ \citenamefont
  {Julienne}}]{schunck_feshbach_2005}%
  \BibitemOpen
  \bibfield  {author} {\bibinfo {author} {\bibfnamefont {C.~H.}\ \bibnamefont
  {Schunck}}, \bibinfo {author} {\bibfnamefont {M.~W.}\ \bibnamefont
  {Zwierlein}}, \bibinfo {author} {\bibfnamefont {C.~A.}\ \bibnamefont {Stan}},
  \bibinfo {author} {\bibfnamefont {S.~M.~F.}\ \bibnamefont {Raupach}},
  \bibinfo {author} {\bibfnamefont {W.}~\bibnamefont {Ketterle}}, \bibinfo
  {author} {\bibfnamefont {A.}~\bibnamefont {Simoni}}, \bibinfo {author}
  {\bibfnamefont {E.}~\bibnamefont {Tiesinga}}, \bibinfo {author}
  {\bibfnamefont {C.~J.}\ \bibnamefont {Williams}}, \ and\ \bibinfo {author}
  {\bibfnamefont {P.~S.}\ \bibnamefont {Julienne}},\ }\href {\doibase
  10.1103/PhysRevA.71.045601} {\bibfield  {journal} {\bibinfo  {journal} {Phys.
  Rev. A}\ }\textbf {\bibinfo {volume} {71}},\ \bibinfo {pages} {045601}
  (\bibinfo {year} {2005})}\BibitemShut {NoStop}%
\bibitem [{\citenamefont {Kotochigova}(2014)}]{kotochigova_controlling_2014}%
  \BibitemOpen
  \bibfield  {author} {\bibinfo {author} {\bibfnamefont {S.}~\bibnamefont
  {Kotochigova}},\ }\href {\doibase 10.1088/0034-4885/77/9/093901} {\bibfield
  {journal} {\bibinfo  {journal} {Rep. Prog. Phys.}\ }\textbf {\bibinfo
  {volume} {77}},\ \bibinfo {pages} {093901} (\bibinfo {year}
  {2014})}\BibitemShut {NoStop}%
\bibitem [{\citenamefont {Kosicki}\ \emph {et~al.}(2020)\citenamefont
  {Kosicki}, \citenamefont {Borkowski},\ and\ \citenamefont
  {{\.Z}uchowski}}]{kosicki_quantum_2020}%
  \BibitemOpen
  \bibfield  {author} {\bibinfo {author} {\bibfnamefont {M.~B.}\ \bibnamefont
  {Kosicki}}, \bibinfo {author} {\bibfnamefont {M.}~\bibnamefont {Borkowski}},
  \ and\ \bibinfo {author} {\bibfnamefont {P.~S.}\ \bibnamefont
  {{\.Z}uchowski}},\ }\href {\doibase 10.1088/1367-2630/ab6c36} {\bibfield
  {journal} {\bibinfo  {journal} {New J. Phys.}\ }\textbf {\bibinfo {volume}
  {22}},\ \bibinfo {pages} {023024} (\bibinfo {year} {2020})}\BibitemShut
  {NoStop}%
\end{thebibliography}
%

\end{document}